\documentstyle[A4,12pt]{article}
\begin{document}
\pagestyle{headings}
\large
\begin{center}
{\Large
{\bf SUPERFIELD FORMULATION OF THE\\
     LAGRANGIAN BRST QUANTIZATION\\ METHOD\\}}
\vspace{.5cm}
P.M.LAVROV\footnote{E-mail: lavrov@tspi.tomsk.su.}\\
{\normalsize\it Tomsk State Pedagogical University, 634041, Russia}\\
P.YU.MOSHIN and A.A.RESHETNYAK\\
{\normalsize\it Tomsk State University, 634050, Russia}
\end{center}
\vspace{.5cm}
\begin{quotation}
 \normalsize
 \setlength{\baselineskip}{15pt}
Lagragian quantization rules for general gauge theories are proposed
on a basis of a superfield formulation of the standard BRST symmetry.
Independence of the $S$-matrix on a choice of the gauge is proved. The
Ward identities in terms of superfields are derived.
\end{quotation}

\section{Introduction}
\hspace*{\parindent}
The majority of field models are based, as a rule, on the fundamental
principle of gauge invariance. On the quantum level it leads to the fact
that there exists a special type of global supersymmetry, i.e. BRST
$\rm{symmetry,}^{1,2}$ underlying the advanced covariant quantization methods
for gauge $\rm{theories.}^{3-5}$

It appears quite natural to refer to papers 4, 5 as the ones to give the
most general form of the corresponding quantization rules. The antisympletic
manifold of the BV $\rm method^{4,5}$ contains the fields $\phi^A$ (including
the initial classical fields, the ghosts, the antighosts and the Lagrangian
multipliers) with assigned to them antifields $\phi^{*}_{A}$ of the
opposite Grassmann parity, the usual sources $J_A$ to the fields $\phi^A$
and finally, the auxiliary fields $\lambda^A$, introducing the gauge to the
theory.

As shown by $\rm{Witten,}^6$ the covariant quantization formalism allows a
geometrical interpretation.

In turn, the Yang--Mills type theories permit one to realize the BRST
symmetry transformations in $\rm{superspace.}^{7-10}$ This being said, the
crucial point of the $\rm formulations^{7-10}$ is the structure of
configuration space of the theories concerned. On the other hand, a closed
form of the Lagrangian quantization rules for general gauge theories that
would enable one to give the BRST transformations a completely geometrical
description has not yet been found.

The purpose of this paper is formulation of the Lagrangian quantization
rules on a basis of a superfield approach, revealing the geometrical
contents of the BRST symmetry.

We use the condensed notations suggested by De $\rm Witt^{11}$; derivatives
with respect to (super)fields are understood as the right-hand and those with
respect to (super-)antifields as the left-hand ones. Left derivatives with
respect to (super)fields are labelled by the subscript $"l"$. The Grassmann
parity of a certain quantity $A$ is denoted $\varepsilon(A)$.

\section{Basic Definitions}
\hspace*{\parindent}
Let us consider superspace $D+1$, parametrized by coordinates
$(x^\mu,\theta)$; $x^\mu$ are the space-time coordinates,
$\mu=(0,1,\ldots,D-1)$; $\theta$ is a scalar Grassmann coordinate.
Let $\Phi^A(\theta)$ be a set of superfields and $\Phi^*_A(\theta)$ be a set
of the corresponding super-antifields
\begin{eqnarray}
 \varepsilon(\Phi^A)\equiv\varepsilon_A,\;\;\;\;\;
 \varepsilon(\Phi^*_A)=\varepsilon_A+1.
\end{eqnarray}

In terms of the superfields and super-antifields we define an antibracket by
the rule
\begin{eqnarray}
 (F,G)&=&\int d\theta\bigg\{\frac{\delta
 F}{\delta\Phi^A(\theta)}\frac{\partial} {\partial\theta}\frac{\delta
 G}{\delta\Phi^*_A(\theta)}(-1)^{\varepsilon_A+1} \nonumber
\\ &&
 -(-1)^{(\varepsilon(F)+1)(\varepsilon(G)+1)}(F\leftrightarrow G)\bigg\}\;,
\end{eqnarray}
where $F=F[\Phi,\Phi^*]$, $G=G[\Phi,\Phi^*]$ are arbitrary functionals
depending on supervariables. From the definition (2) of the antibracket
there follow the properties
\begin{eqnarray}
 && \varepsilon((F,G))=\varepsilon(F)+\varepsilon(G)+1\;,     \nonumber
 \\ &&
 (F,G)=-(G,F)(-1)^{(\varepsilon(F)+1)(\varepsilon(G)+1)}\;,\nonumber
 \\ &&
 (F,GH)=(F,G)H+(F,H)G(-1)^{\varepsilon(G)\varepsilon(H)}\;,\nonumber
 \\ &&
 {((F,G),H)(-1)^{(\varepsilon(F)+1)(\varepsilon(H)+1)}+{\rm cycl.perm.}
 (F,G,H)}\equiv 0\;.
\end{eqnarray}
The last relation is the generalized Jacobi identity for the antibracket.

Let us also introduce operators $\Delta$, $V$ of the form
\begin{eqnarray}
 \Delta=-\int d\theta(-1)^{\varepsilon_A}\frac{\delta_l}{\delta
\Phi^A(\theta)}\frac{\partial}{\partial\theta}\frac{\delta}
 {\delta\Phi^*_A(\theta)}
\end{eqnarray}
\begin{eqnarray}
 V=-\int d\theta\bigg\{\frac{\partial\Phi^*_A(\theta)}{\partial\theta}
 \frac{\delta}{\delta\Phi^*_A(\theta)}+\frac{\partial\Phi^A(\theta)}
 {\partial\theta}\frac{\delta_l}{\delta\Phi^A(\theta)}\bigg\},
\end{eqnarray}
Integration over the coordinate $\theta$ is defined by
\[
  \int d\theta\cdot 1=0\;,\;\;\int d\theta \cdot \theta=1\;,
\]
derivatives with respect to $\theta$ are always undestood as the left-hand
ones.

The algebra of operators (4),(5) has the form
\begin{eqnarray}
 \Delta^2=0\;,\;\;V^2=0\;,\;\;V\Delta+\Delta V=0\;.
\end{eqnarray}
The action of the operators $\Delta$, $V$ upon the antibracket is given
by the following relations
\begin{eqnarray}
 && \Delta (F,G)=(\Delta F,G) -(-1)^{\varepsilon(F)}(F,\Delta G)\;,\nonumber
 \\ &&
 V(F,G)=(VF,G)-(-1)^{\varepsilon(F)}(F,VG)\;.
\end{eqnarray}

\section{Quantization Rules}
\hspace*{\parindent}
Let us now define the generating functional of Green's functions as a
functional depending on the super-antifields $Z=Z[\Phi^*]$ in the form
\begin{eqnarray}
 Z[\Phi^*]=\int
 d\Phi^{\acute{}}\;d\Phi^{*\acute{{}}}\rho[\Phi^{*\acute{{}}}]
 \exp\bigg\{\frac{i}{\hbar}\bigg(S[\Phi^{\acute{{}}},\Phi^{*\acute{{}}}]
 -V^{\acute{{}}}\Psi[\Phi^{\acute{{}}}]-\Phi^*\Phi^{\acute{{}}}\bigg)\bigg\},
\end{eqnarray}
In (8) $S=S[\Phi,\Phi^*]$ is a quantum action satisfying the generating
equation
\begin{eqnarray}
 \frac{1}{2}(S,S)+VS=i\hbar\Delta S
\end{eqnarray}
with the boundary condition
\begin{eqnarray}
 \left.S\right|_{\Phi^*=\hbar=0}=\cal S\;,
\end{eqnarray}
where $\cal S$ is a classical gauge invariant action; $\Psi=\Psi[\Phi]$
is a fermion functional introducing the gauge; $\hbar$ is a Plank constant.
Besides, the following notations
\begin{eqnarray}
 \rho[\Phi^*]=\delta\bigg(\int d\theta\Phi^*(\theta)\bigg),\;\;\;\;\;
 \Phi^*\Phi=\int d\theta\Phi^*_A(\theta)\Phi^A(\theta)
\end{eqnarray}
are used.

An important property of the integrand in (8) for $\Phi^*=0$ is its
invariance under the following global supersymmetry transformations with
a constant Grassmann parameter $\mu$
\begin{eqnarray}
 && \delta\Phi^A(\theta)=\mu\frac{\partial}{\partial\theta}\Phi^A(\theta)\;,
 \nonumber \\ &&
 \delta\Phi^*_A(\theta)=\mu\frac{\partial}{\partial\theta}\Phi^*_A(\theta)+
 \mu\frac{\partial}{\partial\theta}\frac{\delta S}{\delta\Phi^A(\theta)}.
\end{eqnarray}
The transformations (12) realize a superfield form of the BRST symmetry
transformations and permit one to establish the fact that the vacuum
functional $Z_\Psi\equiv Z[0]$ is independent on a choice of the gauge.
Indeed, we shall change the gauge by the rule $\Psi\to\Psi+\delta\Psi$.
In the functional integral for $Z_{\Psi+\delta\Psi}$ we make the change of
variables (12) with the parameter $\mu=\mu[\Phi]$. By virtue of Eq.(9),
$Z_{\Psi+\delta\Psi}$ takes on the form
\begin{eqnarray}
 Z_{\Psi+\delta\Psi}&=&\int
 d\Phi\;d\Phi^{*}\rho[\Phi^{*}]
\exp\bigg\{\frac{i}{\hbar}\bigg(S[\Phi,\Phi^{*}]
 -V\Psi[\Phi] \nonumber
\\ &&
 -V\delta\Psi[\Phi] + i\hbar V \mu[\Phi]\bigg)\bigg\}\;.
\end{eqnarray}
Then, choosing for the parameter $\mu$ the functional
\begin{eqnarray}
 \mu=-\frac{i}{\hbar}\delta\Psi\;,
\end{eqnarray}
we find that $Z_{\Psi+\delta\Psi}=Z_\Psi$ and conclude that the $S$-matrix
is gauge independent.

Eq.(12) implies that from the geometrical viewpoint the operator $V$ (5) can
be considered as a generator of translations in superspace. Given this the
transformations (12) take on the form
\begin{eqnarray}
 && \delta\Phi^A(\theta)=\mu V\Phi^A(\theta)\;,
 \nonumber \\ &&
 \delta\Phi^*_A(\theta)=\mu V\Phi^*_A(\theta)+
 \mu \bigg(S,\Phi^{*}_A(\theta)\bigg)\;.
\end{eqnarray}

Another consequence of validity of the transformations (12) are the Ward
identities for the generating functional of Green's functions. In fact,
making in the functional integral (8) the change of variables (12) and
taking the generating equation for $S=S[\Phi,\Phi^*]$ into account, we
arrive at the relation
\begin{eqnarray}
 && \int
 d\Phi^{\acute{}}\;d\Phi^{*\acute{{}}}\rho[\Phi^{*\acute{{}}}]
 \int d\theta \frac{\partial\Phi^*_A(\theta)}{\partial\theta}
 \Phi^{\acute{{}}A}(\theta)
\exp\bigg\{\frac{i}{\hbar}\bigg(S[\Phi^{\acute{{}}},\Phi^{*\acute{{}}}]
 \nonumber
 \\ &&
\qquad
 -V^{\acute{{}}}\Psi[\Phi^{\acute{{}}}]-\Phi^*\Phi^{\acute{{}}}\bigg)\bigg\}
 =0\;,
\end{eqnarray}
representable, with allowance made for Eq.(8), in the form
\begin{eqnarray}
 -\int d\theta\frac{\partial\Phi^*_A(\theta)}{\partial\theta}\frac{\delta}
 {\delta\Phi^*_A(\theta)}Z[\Phi^*]=VZ[\Phi^*]=0\;.
\end{eqnarray}
Now, define the generating functional of vertex functions (effective
action) depending on the superfields $\Gamma=\Gamma[\Phi]$ by the Legendre
transformation for $\ln Z$ with respect to the super-antifields $\Phi^*$
\begin{eqnarray}
 \Gamma[\Phi]=\frac{\hbar}{i}\ln Z[\Phi^*]+\Phi^*\Phi,\;\;
 \Phi^A(\theta)=-\frac{\hbar}{i}\frac{\delta}{\delta\Phi^*_A(\theta)}
 \ln Z[\Phi^*]\;,
\end{eqnarray}
then the identity (17) can be represented in the form
\begin{eqnarray}
 -\int d\theta\frac{\partial\Phi^A(\theta)}{\partial\theta}\frac{\delta_l}
 {\delta\Phi^A(\theta)}\Gamma[\Phi]=V\Gamma[\Phi]=0\;.
\end{eqnarray}
Geometrically, the Ward identities (17), (19) imply the fact that the
functionals $Z[\Phi^*]$, $\Gamma[\Phi]$ are invariant under supertranslations
with respect to the coordinate $\theta$.

\section{Relation to the BV Quantization Scheme}
\hspace*{\parindent}
It appears very important to establish a relation between the superfield
approach in question and the BV quantization rules. To this end, note that
the components of superfields $\Phi^A(\theta)$ and super-antifields
$\Phi^*_A(\theta)$ are defined by expansions in $\theta$
\begin{eqnarray}
 && \Phi^A(\theta)=\phi^A+\lambda^A\theta\;,\;\;
 \Phi^*_A(\theta)=\phi^*_A-\theta J_A\;,  \nonumber
 \\ &&
 \varepsilon(\phi^A)=\varepsilon(J_A)=\varepsilon_A\;,\;\;
 \varepsilon(\phi^*_A)=\varepsilon(\lambda^A)=\varepsilon_A+1
\end{eqnarray}
and coincide with the set of variables in the BV quantization scheme (the
choice of signs in Eq. (20) is due to considerations of convinience).

Consider by virtue of Eq. (20) the component form of the basic definitions
and relations given above.

First, the antibracket (2) is representable in terms of the component
fields $\phi^A,\;\phi^*_A,\;\lambda^A,\;J_A$ as follows
\begin{eqnarray}
 (F,G)=\frac{\delta F}{\delta\phi^A} \frac{\delta G}{\delta \phi^*_A}
 -(-1)^{(\varepsilon(F)+1)(\varepsilon(G)+1)}(F\leftrightarrow G)\;.
\end{eqnarray}
Eq.(21) coincides with the usual definition of the antibracket in the
framework of BV quantization method.

Second, the corresponding component expressions for the operators $\Delta$,
$V$ (4), (5) read
\begin{eqnarray}
 \Delta &=& (-1)^{\varepsilon_A}\frac{\delta_l}{\delta \phi^A} \frac
 {\delta}{\delta \phi^*_A}\;,
\end{eqnarray}
\begin{eqnarray}
 V=-J_A \frac{\delta}{\delta \phi^*_A}-(-1)^{\varepsilon_A}\lambda^A
 \frac{\delta_l}{\delta \phi^A}\;.
\end{eqnarray}
\hspace*{\parindent}
 In virtue of Eqs. (21), (23) we find that the transformations (12) take on
 the form
\begin{eqnarray}
 \delta\phi^A=\lambda^A\mu,\;\;\;\;\delta \lambda^A=0,\nonumber\\
\end{eqnarray}
\[
 \delta\phi^*_A=\mu\bigg(\frac{\delta S}{\delta\phi^A}-J_A\bigg),\;\;\;\;
 \delta J_A=0.
\]
 Next, making use of Eq. (23), one readily obtains the component form of the
 Ward identities (17), (19) for the functionals $Z(\phi^*,J)\equiv Z[\phi^*]$,
 $\Gamma(\phi,\lambda)\equiv\Gamma[\Phi]$
\begin{eqnarray}
 J_A\frac{\delta}{\phi^*_A}Z(\phi^*,J)=0,\;\;\;\;
 \lambda^A \frac{\delta}{\delta \phi^A} \Gamma(\phi,\lambda)=0,
\end{eqnarray}
 Finally, the integration measure in Eq. (8) is understood as follows
\begin{eqnarray}
 d\Phi\;d\Phi^*\;\rho(\Phi^*)=d\phi\;d\phi^*\;d\lambda\;dJ\;\delta(J)
\end{eqnarray}
 and the functional $\Phi^*\Phi$ has the form
\begin{eqnarray}
 \Phi^*_A \Phi^A = \phi^*_A \lambda^A - J_A\phi^A\;.
\end{eqnarray}
\hspace*{\parindent}
 All things considered, choosing, by virtue of Eqs. (21), (22), (23), for a
 solution of the generating equation (9) with the boundary condition (10) a
 functoinal $S=S[\Phi,\Phi^*]$ such that
\begin{eqnarray}
 \left.S[\Phi,\Phi^*]\right|_{J=0}=\overline{S}(\phi,\phi^*)+\phi^*_A
 \lambda^A,
\end{eqnarray}
 where $\overline{S}$ satisfies the usual master equation of Refs. 4, 5
\begin{eqnarray}
 \frac{1}{2}(\overline{S},\,\overline{S})=i\hbar\Delta\overline{S},\;\;\;\;\;
 \left.\overline{S}\right|_{\phi^*=\hbar=0}=\cal S,
\end{eqnarray}
 we arrive, making use of Eqs. (26), (27) at the following
 representation for the generating functional of Green's functions
 $Z=Z(J)$ of the fields $\phi^A$
\begin{eqnarray}
 Z(J)&=&\left.Z[\Phi^*]\right|_{\phi^*=0}=\int
 d\phi\;d\phi^*\;d\lambda\exp\bigg\{\frac{i}{\hbar}\bigg[\overline{S}
 (\phi,\phi^*)\nonumber\\ &&
 +\bigg(\phi^*_A-\frac{\delta\Psi}
 {\delta\phi^A}\bigg)\lambda^A+J_A\phi^A\bigg]\bigg\}.
\end{eqnarray}
 The above relation defines, with allowance made for Eq. (29), the generating
 functional of Green's functions in the framework of the BV quantization
 formalism.

\section{Conclusion}
\hspace*{\parindent}
 In this paper the Lagrangian quantization rules for general gauge theories
 on a basis of a superfield realization of the standard BRST symmetry are
 presented. The $S$-matrix is shown to be gauge independent. The Ward
 identities (17), (19), corresponding to the superfield form (12) (or,
 equivalently, (15)) of the BRST transformations, imply invariance of the
 functionals $Z[\Phi^*]$, $\Gamma[\Phi]$ under translations in superspace
 $(x^\mu,\theta)$ with respect to the Grassmann coordinate $\theta$. It is
 shown that the special choice (28), (29) of a solution to the equation (9)
 determining the boson functional $S$ leads to the generating functional of
 Green's functions of the BV quantization scheme.

\vspace{1cm}
\noindent{\bf Acknoledgment}

\vspace{.5cm}
 The work is supported in part by the International Scientific Foundation,
 grant RI 1300 and by the Russian Foundation for Fundamental Research,
 project No. 94--02--03234.

\newpage
\pagestyle{myheadings}
\markright{}
\begin{center}
\Large\bf References
\end{center}
\begin{itemize}
\item[1.] C.Becchi, A.Rouet, R.Stora, \underline{Commun. Math. Phys.}
 42 (1975) 127.
\item[2.] I.V.Tyutin, preprint Lebedev Inst. No. 39, 1975.
\item[3.] B. de Wit, J.W. van Holten, \underline{Phys. Lett.} B79
 (1978) 389.
\item[4.] I.A.Batalin, G.A.Vilkovisky, \underline{Phys. Lett.} 102B
 (1981) 27.
\item[5.] I.A.Batalin, G.A.Vilkovisky, \underline{Phys. Rev.} D28
 (1983) 2567.
\item[6.] E.Witten, \underline{Mod. Phys. Lett.} A5 (1990) 487.
\item[7.] L.Bonora, M.Tonin, \underline{Phys. Lett.} B98 (1981) 48.
\item[8.] L.Bonora, P.Pasti, M.Tonin, \underline{J. Math. Phys.} 23
 (1982) 839.
\item[9.] C.M.Hull, B.Spence, J.L.Vazquez-Bello, \underline{Nucl. Phys.}
 B348 (1990) 108.
\item[10.] L.Baulieu, \underline{Phys. Rep.} 129 (1985) 1.
\item[11.] B.S. De Witt, \underline{Dynamical Theory of Groups and Fields}
 (Gordon and Breach, New York, 1965).
\end{itemize}
\end{document}